\documentclass[12pt]{article}
\usepackage{amsmath,amssymb}
\usepackage{graphicx}
\usepackage{wrapfig}

\begin{document}

\title{$\eta$ and $\eta'$ photoproduction with EtaMAID including Regge phenomenology}


\author{V.L. Kashevarov, L. Tiator, M. Ostrick\\
Institut f\"ur Kernphysik, Johannes Gutenberg-Universit\"at\\
D-55099 Mainz, Germany}

\maketitle

\begin{abstract}

We present a new version of the EtaMAID model for $\eta$ and $\eta'$
photoproduction on nucleons. The model includes 23 nucleon
resonances parameterized with Breit-Wigner shapes. The background is
described by vector and axial-vector meson exchanges in the $t$
channel using the Regge cut phenomenology. Parameters of the
resonances were obtained from a fit to available experimental data
for $\eta$ and $\eta'$ photoproduction on protons and neutrons. The
nature of the most interesting observations in the data is
discussed.
\end{abstract}

EtaMAID is an isobar model~\cite{MAID, MAIDr} for $\eta$ and $\eta'$ photo-
and electroproduction on nucleons. The model includes a non-resonant
background, which consists of nucleon Born terms in the $s$ and $u$
channels and the vector meson exchange in the $t$ channel, and
$s$-channel resonance excitations, parameterized by Breit-Wigner
functions with energy dependent widths. The EtaMAID-2003 version
describes the experimental data available in 2002 reasonably well,
but fails to reproduce the newer polarization data obtained in
Mainz~\cite{TF_MAMI}. During the last two years the EtaMAID model
was updated~\cite{MAID15, MAID16, MAID17} to describe the new data
for $\eta$ and $\eta'$ photoproduction on the proton. The presented
EtaMAID version includes also $\eta$ and $\eta'$ photoproduction on
the neutron.

At high energies, $W > 3$ GeV, Regge cut phenomenology was applied.
The models include $t$-channel exchanges of vector ($\rho$ and
$\omega$) and axial vector ($b_1$ and $h_1$) mesons as Regge
trajectories. In addition to the Regge trajectories, also Regge cuts
from rescattering $\rho{\mathbb P}$, $\rho f_2$ and $\omega{\mathbb
P}$, $\omega f_2$ were added, where ${\mathbb P}$ is the Pomeron
with quantum numbers of the vacuum $0^+(0^{++})$ and $f_2$ is a
tensor meson with quantum numbers $0^+(2^{++})$. The obtained
solution describes the data up to $E_{\gamma}=8$~GeV very well. For
more details see Ref.~\cite{Regge}. Energies below $W=2.5$~GeV are
dominated by nucleon resonances in the $s$ channel. All known
resonances with an overall rating of two stars and more were
included in the fit. To avoid double counting from $s$ and $t$
channels in the resonance region, low partial waves with L up to 4
were subtracted from the $t$-channel Regge contribution.

\begin{figure}
\begin{center}
\includegraphics[width=0.6\textwidth]{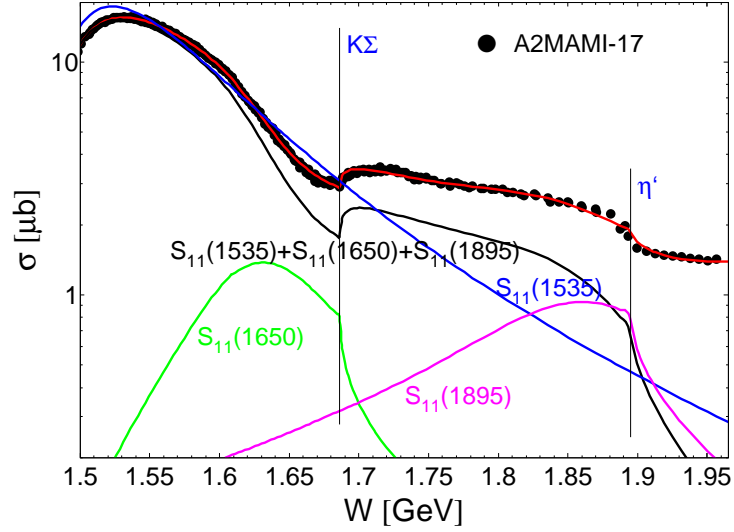}
\caption{Total cross section of the $\gamma p \to \eta p$ reaction
with partial contributions of the main nucleon resonances. Red line:
New EtaMAID solution. Vertical lines correspond to thresholds of
$K\Sigma$ and $\eta' N$ photoproduction. Data:
A2MAMI-17~\cite{MAID17}. } \label{fig1}
\end{center}
\end{figure}
\begin{figure}
\begin{center}
\includegraphics[width=0.6\textwidth]{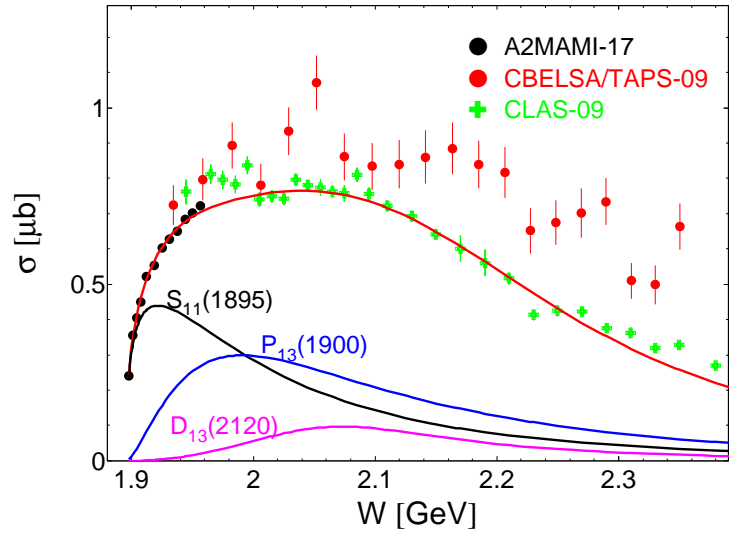}
\caption{Total cross section of the $\gamma p \to \eta' p$ reaction
with partial contributions of the main nucleon resonances. Red line:
New EtaMAID solution. Data: A2MAMI-17~\cite{MAID17},
CBELSA/TAPS-09~\cite{dcs_ELSA}, and CLAS-09~~\cite{dcs_CLAS}. }
\label{fig2}
\end{center}
\end{figure}
The most interesting fit results are presented in Figs. 1-5 together
with corresponding experimental data.

In Fig.~\ref{fig1}, the total $\gamma p \to \eta p$ cross section is
shown. A key role in the description of the investigated reactions
is played by three $s$-wave resonances N(1535)$1/2^-$,
N(1650)$1/2^-$, and N(1895)$1/2^-$, see partial contributions of
these resonances in Fig.~\ref{fig1}. The first two give the main
contribution to the total cross section and are known very well. An
interference  of these two resonances is mainly responsible for the
dip at $W=1.68$~GeV. However, the narrowness of this dip we explain
as a threshold effect due to the opening of the K$\Sigma$ decay
channel of the $N(1650)1/2^-$ resonance. The third one,
N(1895)$1/2^-$, has only a 2-star overall status according to the
PDG review~\cite{PDG}. But we have found that namely this resonance
is responsible for the cusp effect at $W=1.96$~GeV (see magenta line
in Fig.~\ref{fig1}) and provides a fast increase of the total cross
section in the $\gamma p \to \eta' p$ reaction near threshold (see
black line in Fig.~\ref{fig2}). A good agreement with the
experimental data was obtained for the cross sections of the $\gamma
p \to \eta' p$ reaction, Fig.~\ref{fig2}. The main contributions to
this reaction come from $N(1895)1/2^-$, $N(1900)3/2^+$, and
$N(2130)3/2^-$ resonances.

\begin{figure}
\begin{center}
\includegraphics[width=0.6\textwidth]{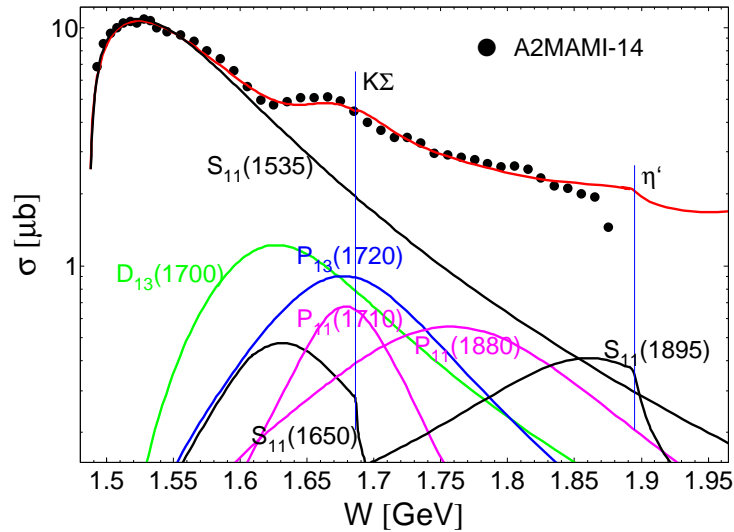}
\caption{Total cross section of the $\gamma n \to \eta n$ reaction
with partial contributions of the main nucleon resonances. Red line:
New EtaMAID solution. Data: A2MAMI-14~\cite{MAMI-14}. } \label{fig3}
\end{center}
\end{figure}
Very interesting results were obtained during the last few years for
the $\gamma n \to \eta n$ reaction. The excitation function for this
reaction shows an unexpected narrow structure at $W \sim 1.68$~GeV,
which is not observed in $\gamma p \to \eta p$. As an example, the
total cross section measured with highest statistics in Mainz
\cite{MAMI-14} is shown in Fig.~\ref{fig3}. The nature of the narrow
structure has been explained by different authors as a new exotic
nucleon resonance, or a contribution of intermediate strangeness
loops, or interference effects of known nucleon resonances, see
Ref.~\cite{Krusche}. In our analyses, the narrow structure is
explained as the interference of $s$, $p$, and $d$ waves, see
partial contributions of the resonances in Fig.~\ref{fig3}. Our full
solution, red line in Fig.~\ref{fig3}, describes the data up to $W
\sim 1.85$~GeV reasonably well and shows a cusp-like structure at $W
= 1.896$~GeV similar as in Fig.~\ref{fig2} for the $\gamma p \to
\eta p$ reaction. However, the data demonstrate a cusp-like effect
at the energy of $\sim 50$ MeV below. This remains an open question
for our analyses as well as for the final state effects in the data
analysis.

\begin{figure}
\begin{center}
\resizebox{1.0\textwidth}{!}{\includegraphics{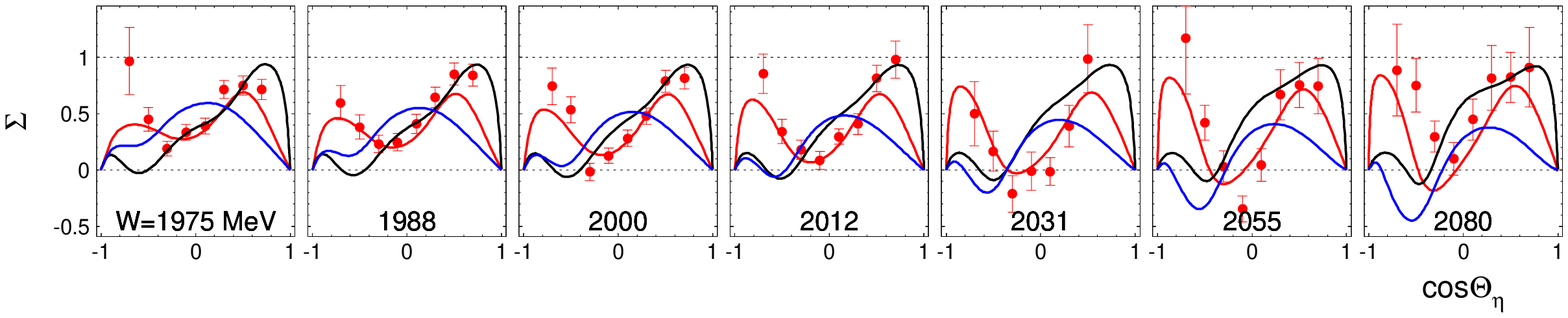}}
\caption{Beam asymmetry $\Sigma$ for the $\gamma p \to \eta p$
reaction. Red line: New EtaMAID solution. Results of the refit to
the data without $N(2120)3/2^-$ are shown by the black lines and
without $N(2060)5/2^-$ - blue lines. Data: CLAS-17
\cite{Sigma_CLAS-17}, } \label{fig4}
\end{center}
\end{figure}
Recently, the CLAS collaboration reported a measurement of the beam
asymmetry $\Sigma$ for both $\gamma p \to \eta p$ and $\gamma p \to
\eta' p$ reactions \cite{Sigma_CLAS-17}. At high energies, $W >
2$~GeV, the $\gamma p \to \eta p$ data have maximal $\Sigma$
asymmetry at forward and backward directions, see Fig.~\ref{fig4}.
We have found that an interference of $N(2120)3/2^-$ and
$N(2060)5/2^-$ resonances is responsible for such an angular
dependence. The data was refitted excluding the resonances with mass
around 2 GeV. The most significant effect we have found by refitting
without $N(2120)3/2^-$ (black line) and $N(2060)5/2^-$ (blue line).
The red line is our full solution.

\begin{figure}
\begin{center}
\resizebox{0.7\textwidth}{!}{\includegraphics{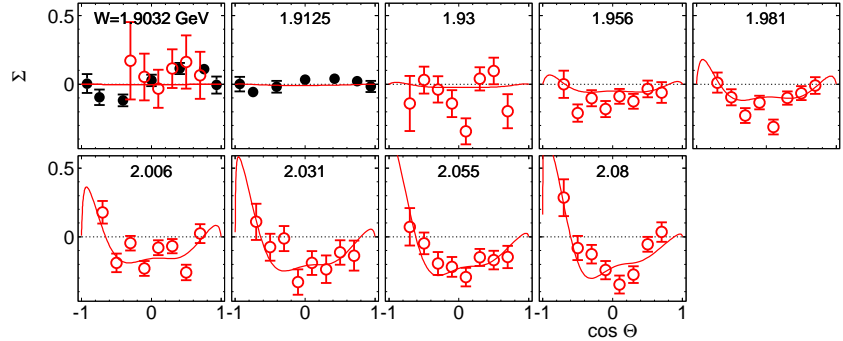}}
\caption{Beam asymmetry $\Sigma$ for the $\gamma p \to \eta' p$
reaction. Red line: New EtaMAID solution. Data: GRAAL-15
\cite{GRAAL-15} (black), CLAS-17 \cite{Sigma_CLAS-17} (red). }
\label{fig5}
\end{center}
\end{figure}
The beam asymmetry $\Sigma$ for $\gamma p \to \eta' p$ reaction is
presented in Fig.~\ref{fig5} with the GRAAL data \cite{GRAAL-15}
having a nodal structure near threshold. Such a shape of the angular
dependence could be explained by interference of $s$ and $f$ or $p$
and $d$ waves. However, the energy dependence is inverted in all
models. The EtaMAID-2016 solution \cite{MAID16} describes the shape
of the GRAAL data for $\Sigma$, but not the magnitude. The new CLAS
data \cite{Sigma_CLAS-17} can not solve this problem because of poor
statistics new threshold. Our new solution describes the $\Sigma$
data well at $W >1.95$~GeV.

In summary, we have presented a new version $\eta$MAID-2017n updated
with new resonances and new experimental data. The model describes
the data currently available for both $\eta$ and $\eta'$
photoproduction on protons and neutrons. The cusp in the $\eta$
total cross section, in connection with the steep rise of the
$\eta'$ total cross section from its threshold, is explained by a
strong coupling of the $N(1895)1/2^-$ to both channels. The narrow
bump in $\eta n$ and the dip in $\eta p$ channels have a different
origin: the first is a result of an interference of a few
resonances, and the second is a threshold effect due to the opening
of the K$\Sigma$ decay channel of the $N(1650)1/2^-$ resonance. The
angular dependence of $\Sigma$ for $\gamma p \to \eta p$ at
$W>2$~GeV is explained by an interference of $N(2120)3/2^-$ and
$N(2060)5/2^-$ resonances. The near threshold behavior of $\Sigma$
for $\gamma p \to \eta' p$, as seen in the GRAAL data, is still an
open question. A further improvement of our analysis will be
possible with additional polarization observables which soon should
come from the A2MAMI, CBELSA/TAPS, and CLAS collaborations.

This work was supported by the Deutsche Forschungsgemeinschaft (SFB 1044).


\end{document}